\documentclass[conference,hidelinks,10pt]{IEEEtran}

\ifCLASSINFOpdf
\else
\usepackage[dvips]{graphicx}
\fi
\usepackage{url}

\usepackage{graphicx} 
\usepackage{booktabs} 
\usepackage{mathrsfs}
\usepackage{amsmath}
\usepackage{amssymb}
\usepackage{bm}
\usepackage{mathtools}
\usepackage{cite}
\usepackage[acronym,shortcuts]{glossaries}
\usepackage{enumitem} 
\usepackage{comment}
\usepackage{algorithm, algpseudocode}
\usepackage{mathtools}
\usepackage{lipsum}
\usepackage{mleftright}
\usepackage{orcidlink}
\usepackage{soul}       

\setlist[itemize]{noitemsep} 

\DeclareMathOperator*{\argmin}{arg\,min}

\providecommand{\UL}{\mathrel{\raise-4pt\hbox{\hglue -2.8ex
\vrule height .1ex width 2.3ex
\vrule height 3ex width .1ex
\hglue .4ex}}}
\providecommand{\ul}{\mathrel{\raise-2pt\hbox{\hglue -2.3ex
\vrule height .1ex width 2ex
\vrule height 2ex width .1ex
\hglue .4ex}}}
\providecommand{\ub}{
\mathrel{\hbox{\hglue -2.8ex \vrule height 2ex width .06ex}
\raise2ex\hbox{\hglue -0.1ex\vrule height .1ex width 2.5ex}
\hbox{\hglue -0.1ex \vrule height 2ex width .1ex
\hglue .4ex}}}
\providecommand{\lb}{
\mathrel{\raise-0.5ex
\hbox{\hglue -2.8ex \vrule height 2ex width .1ex
\vrule height .1ex width 2.5ex
\vrule height 2ex width .1ex
\hglue .4ex}}}
\providecommand{\UB}{
\mathrel{
\raise-1ex\hbox{\hglue -1.8em \vrule height 3ex width .1ex}
\raise2ex\hbox{\hglue -0.1ex\vrule height .1ex width 1.5em}
\raise-1ex\hbox{\hglue -0.1ex \vrule height 3ex width .1ex
\hglue .4ex}}}
\providecommand{\LB}{
\mathrel{\raise-1ex
\hbox{\hglue -1.8em \vrule height 3ex width .1ex
\vrule height .1ex width 1.5em
\vrule height 3ex width .1ex
\hglue .4ex}}}
\usepackage{trimclip}
\newif\iflclip
\newif\ifbclip
\newif\ifrclip
\newif\iftclip
\def\CLIP{\dimexpr\fboxrule+.2pt\relax}
\def\nulclip{0pt}
\newcommand\partbox[2]{%
\lclipfalse\bclipfalse\rclipfalse\tclipfalse%
\let\lkern\relax\let\rkern\relax%
\let\lclip\nulclip\let\bclip\nulclip\let\rclip\nulclip\let\tclip\nulclip%
\parseclip#1\relax\relax%
\iflclip\def\lkern{\kern\CLIP}\def\lclip{\CLIP}\fi
\ifbclip\def\bclip{\CLIP}\fi
\ifrclip\def\rkern{\kern\CLIP}\def\rclip{\CLIP}\fi
\iftclip\def\tclip{\CLIP}\fi
\lkern\clipbox{\lclip{} \bclip{} \rclip{} \tclip}{\fbox{#2}}\rkern%
}
\def\parseclip#1#2\relax{%
\ifx l#1\lcliptrue\else
\ifx b#1\bcliptrue\else
\ifx r#1\rcliptrue\else
\ifx t#1\tcliptrue\else
\fi\fi\fi\fi
\ifx\relax#2\relax\else\parseclip#2\relax\fi
}
\usepackage{efbox}

\newacronym{SVD}{SVD}{singular value decomposition}
\newacronym{DCM}{DCM}{double-centering matrix}
\newacronym{3D}{3D}{three-dimensional}
\newacronym{GA}{GA}{genie-aided}
\newacronym{EA}{EA}{``\emph{estimate-then-average}''}
\newacronym{AE}{AE}{``\emph{average-then-estimate}''}
\newacronym{IRS}{IRS}{intelligent reflecting surface}
\newacronym{RSSI}{RSSI}{received signal strength indicator}
\newacronym{SotA}{SotA}{state-of-the-art}
\newacronym{CSI}{CSI}{channel state information}
\newacronym{D2D}{D2D}{device-to-device}
\newacronym{RR}{RR}{round-robin}
\newacronym{DA}{DA}{Dutch auction}
\newacronym{AV}{AV}{autonomous vehicle}
\newacronym{CWFL}{CWFL}{clustered WFL}
\newacronym{WFL}{WFL}{wireless federated learning}
\newacronym{RSMA}{RSMA}{rate splitting multiple access}
\newacronym{IoT}{IoT}{Internet-of-Things}
\newacronym{TDMA}{TDMA}{time-domain multiple access}
\newacronym{NOMA}{NOMA}{non-orthogonal multiple access}
\newacronym{ML}{ML}{machine learning}
\newacronym{MIMO}{MIMO}{multiple-input multiple-output}
\newacronym{CT}{CT}{compute-then-transmit}
\newacronym{FP}{FP}{fractional programming}
\newacronym{CF-mMIMO}{CF-mMIMO}{cell free massive MIMO}
\newacronym{iid}{i.i.d.}{independent and identically distributed}
\newacronym{AD}{AD}{autonomous driving}
\newacronym{DL}{DL}{downlink}
\newacronym{UL}{UL}{uplink}
\newacronym{IC}{IC}{interference cancellation}
\newacronym{SIC}{SIC}{successive interference cancellation}
\newacronym{BS}{BS}{base station}
\newacronym{TX}{TX}{transmit}
\newacronym{RX}{RX}{receive}
\newacronym{MU}{MU}{multi-user}
\newacronym{SISO}{SISO}{single-input single-output}
\newacronym{AWGN}{AWGN}{additive white Gaussian noise}
\newacronym{SINR}{SINR}{signal-to-interference-and-noise ratio}
\newacronym{FL}{FL}{federated learning}
\newacronym{CPU}{CPU}{central processing unit}
\newacronym{KNN}{KNN}{K-nearest-neighbor}
\newacronym{RF}{RF}{radio frequency}
\newacronym{GD}{GD}{gradient descent}
\newacronym{V2X}{V2X}{vehicle-to-anything}

\newacronym{RSS}{RSS}{received signal strength}
\newacronym{FIM}{FIM}{fisher information matrix}
\newacronym{ToA}{ToA}{time of arrival}
\newacronym{ToF}{ToF}{time of flightl}
\newacronym{AoA}{AoA}{angle of arrival}
\newacronym{GP}{GP}{Gaussian process}
\newacronym{2D}{2D}{two-dimensional}
\newacronym{GPR}{GPR}{Gaussian process regression}
\newacronym{GNSS}{GNSS}{global navigation satellite systems}
\newacronym{B5G}{B5G}{beyond fifth-generation}
\newacronym{RRH}{RRH}{remote radio head}
\newacronym{GPS}{GPS}{global positioning system}
\newacronym{RFID}{RFID}{radio frequency identification}
\newacronym{TCAS}{TCAS}{traffic alert and collision avoidance systems}
\newacronym{RMSE}{RMSE}{root mean square error}
\newacronym{SGD}{SGD}{stochastic gradient descent}
\newacronym{PDF}{PDF}{probability density function}
\newacronym{CU}{CU}{computing unit}
\newacronym{DM-MIMO}{DM-MIMO}{distributed massive multiple-input multiple-output}
\newacronym{LOS}{LOS}{line-of-sight}
\newacronym{NLOS}{NLOS}{non-line-of-sight}
\newacronym{ROI}{ROI}{region of interest}
\newacronym{AP}{AP}{access point}
\newacronym{TDOA}{TDOA}{time difference of arrival}
\newacronym{UE}{UE}{user equipment}
\newacronym{dB}{dB}{decibel}
\newacronym{RIS}{RIS}{reconfigurable intelligent surface}
\newacronym{CG}{CG}{conjugate gradient}

\newacronym{PG}{PG}{proximal gradient}
\newacronym{SVT}{SVT}{singular value thresholding}
\newacronym{NN}{NN}{nuclear norm}
\newacronym{NMSE}{NMSE}{normalized mean square error}
\newacronym{MC}{MC}{matrix completion}
\newacronym{NP}{NP}{non-deterministic polynomial-time}
\newacronym{EDM}{EDM}{euclidean distance matrix}
\newacronym{SC}{SC}{soft-connected}
\newacronym{CRLB}{CRLB}{Cramér-Rao Lower Bound}
\newacronym{PoA}{PoA}{phase of arrival}
\newacronym{UAV}{UAV}{unmanned aerial vehicle}
\newacronym{VR}{VR}{virtual reality}
\newacronym{MDS}{MDS}{multidimensional scaling}

\newacronym{RBL}{RBL}{rigid body localization}
\newacronym{RBT}{RBT}{rigid body tracking}
\newacronym{SC-RBL}{SC-RBL}{soft-connected RBL}
\newacronym{W-RBL}{W-RBL}{\underline{wireless} RBL}

\newacronym{SDP}{SDP}{semidefinite programming}
\newacronym{JCAS}{JCAS}{joint communication and sensing}
\newacronym{SDR}{SDR}{semi-definite relaxation}

\newacronym{OPP}{OPP}{orthogonal Procrustes problem}
\newacronym{SLAM}{SLAM}{simultaneous localization and mapping}
\newacronym{WLS}{WLS}{weighted least square}
\newacronym{SI}{SI}{soft-impute}

\newacronym{CSDP}{CSDP}{constrained semidefinite programming}
\newacronym{iff}{iff}{if and only if}
\newacronym{UDPN}{UDPN}{unit disk planar network}
\newacronym{BFS}{BFS}{breadth first search}
\newacronym{AODV}{AODV}{ad hoc on-demand distance vector}
\begin{document}

\title{A Robust Routing Protocol for 5G Mesh Networks}

\author{\IEEEauthorblockN{Niclas~F\"uhrling\textsuperscript{\orcidlink{0000-0003-1942-8691}}, Ivan Alexander Morales Sandoval\textsuperscript{\orcidlink{0000-0002-8601-5451}}and Giuseppe Thadeu Freitas de Abreu\textsuperscript{\orcidlink{0000-0002-5018-8174}}}
\IEEEauthorblockA{\textit{School of Computer Science and Engineering,
Constructor University, Bremen, Germany}\\ 
{\small\tt [nfuehrling, imorales, gabreu]@constructor.university}\\[-3ex]
}
}

\setlength{\parskip}{0pt}

\maketitle

\begin{abstract}
We consider a novel routing protocol suitable for ad-hoc networks with dynamically changing topologies, such as DECT 2020 NR (NR+) systems, which often lead to missing links between the nodes and thus, incomplete or inefficient routes.
A key point of the proposed protocol is the combination of network discovery and matrix completion techniques, which allow the nodes to establish communication paths efficiently and reliably.
Additionally, multihop localization is performed to estimate the location of the nodes without needing to broadcast each node's geographical position, thus preserving privacy during the routing process and enabling nodes in the network to independently find potentially missing paths in a decentralized manner instead of flooding the whole network.
Simulation results illustrate the good performance of the proposed technique in terms of the average number of hops of the obtained routes in different scenarios, with different network densities and amounts of incompleteness.
\end{abstract}


\IEEEpeerreviewmaketitle

\vspace{-1ex}
\section{Introduction}

\IEEEPARstart{N}{etwork} routing is an important research topic in the field of wireless communication systems \cite{Alotaibi_2012, Altayeb_2013, Zagrouba_2021}, especially in the context of ad-hoc and mesh networks, where nodes might be mobile and the network topology dynamic.
In such networks, routing protocols play a crucial role in establishing communication paths between the nodes, ensuring that the data packets are delivered efficiently and reliably \cite{Paul_2011, Grohmann_2020, Altenhofen_2024}.
Recently, the DECT 2020 NR\footnote{Other dynamic ad-hoc mesh network technologies that currently gain momentum are, LoRaHop \cite{Tian_2023} and Bluetooth Low Energy (BLE) mesh networks \cite{Ghori_2020}, which are also suitable for the proposed routing protocol.}\cite{etsi2020dect,Kovalchukov_2022,Haque_2024} standard was released by the European Telecommunications Standards Institute (ETSI), making it the first non-cellular 5G technology with peer-to-peer mesh networking capabilities.
While the parameters and protocols on the physical layer are well-defined within the standard, the routing protocols are still an open research area, since it is up to the user to decide which routing protocol to use.
Thus, the need for advanced routing protocols that can handle the unique characteristics of these emerging systems and the applications they enable is now apparent.

Research on routing protocols has been conducted for a long time, with one of the early examples for ad-hoc networks being the \ac{AODV} routing protocol \cite{Perkins_1999}.
However, novel approaches are still being developed to address the unique challenges of modern wireless communication systems, in particular opportunistic routing algorithms \cite{Boukerche_2014} which exploit the broadcast nature of the wireless medium to improve the reliability of the communication links.

The main challenges faced by \ac{SotA} routing protocols targeting modern communication systems are their reliability, resilience, and more recently privacy and security.
In response to these challenges, the original \ac{AODV} algorithm has been extended to better accommodate the requirements of wireless multi-hop mobile ad-hoc networks (MANETs) \cite{Paul_2011} via static routers and router count metrics.
In turn, network resilience was improved in \cite{Grohmann_2020} through dynamic routing and the integration of disjoint paths, and an opportunistic method for resilient routing in highly dynamic mesh networks using network coding was proposed in \cite{Altenhofen_2024}.

A critical flaw of on-demand routing protocols is, however, that they cannot dynamically adapt to new links which were originally missing, since routing tables are kept frozen for a certain period of time after discovery, potentially leading to suboptimal routes between the nodes.
In view of the latter, we propose that the issue of missing links in a network be solved via a matrix completion approach.

Many matrix completion techniques have been proposed in the literature \cite{Chen_2022, Nguyen2019, OptSpace, Yao_2019} as a way to estimate the missing entries of a matrix based on the knowledge of the observed entries.
However, \ac{SotA} methods share a common assumption that the entries of the incomplete matrix can take on any number in the real $\mathbb{R}$ or complex sets $\mathbb{C}$, which is not suitable to the routing problem since the number of hops in a route must be integer.
A recent variation, suitable for the proposed scenario, is the discrete-aware matrix completion scheme proposed in \cite{Iimori_2020, Nic_Asilo_2024}.
Additionally, multihop localization techniques have been proposed to estimate the location of nodes in a network based only on the number of hops connecting them, which can be used to find the shortest path between any two nodes \cite{Rahmatollahi_2011, Rahmatollahi_2012} without the need to share their position nor their distance with neighboring nodes, thus preserving privacy.

Given all the above, we propose in this article a novel robust routing protocol that is suitable for DECT 2020 NR 5G mesh networks \cite{etsi2020dect ,Kovalchukov_2022, Haque_2024}, which can handle the dynamic topologies and potential missing links between the nodes, by first performing discrete-aware matrix completion to estimate the missing hops between the nodes; followed by a multihop localization to estimate the location of the nodes;
and finally, local route discovery performed to find the shortest path between the nodes, which is based on the estimated location of the nodes and the known radio range of the nodes.
Overall, the proposed routing protocol can establish communication paths between pairs of nodes efficiently and reliably, by enabling a decentralized routing process iteration that avoids flooding the whole network with routing messages, in addition to estimating the locations of the nodes, while preserving privacy.

The structure of the remainder of the article is as follows.
First, a description of the system and network model is offered in Section \ref{sec:prior}.
Then, in Section \ref{sec:prop}, after a short introduction to the \ac{SotA} routing approach for partially connected networks is given, the proposed method to find the missing entries of the incomplete hop matrix is introduced, followed by the multihop localization scheme, and the local route discovery.
Finally, a comparison of the proposed scheme with the conventional \ac{BFS} algorithm performing network discovery on the network with partially missing edges, is offered in Section \ref{sec:res}, followed by further performance evaluation in terms of the average number of hops for various network densities and incompleteness levels.

\vspace{-1ex}
\section{Network Model}
\label{sec:prior}
\vspace{-1ex}

\subsection{System Model}
\vspace{-1ex}

\begin{figure}[H]
\centering
\includegraphics[width=\columnwidth]{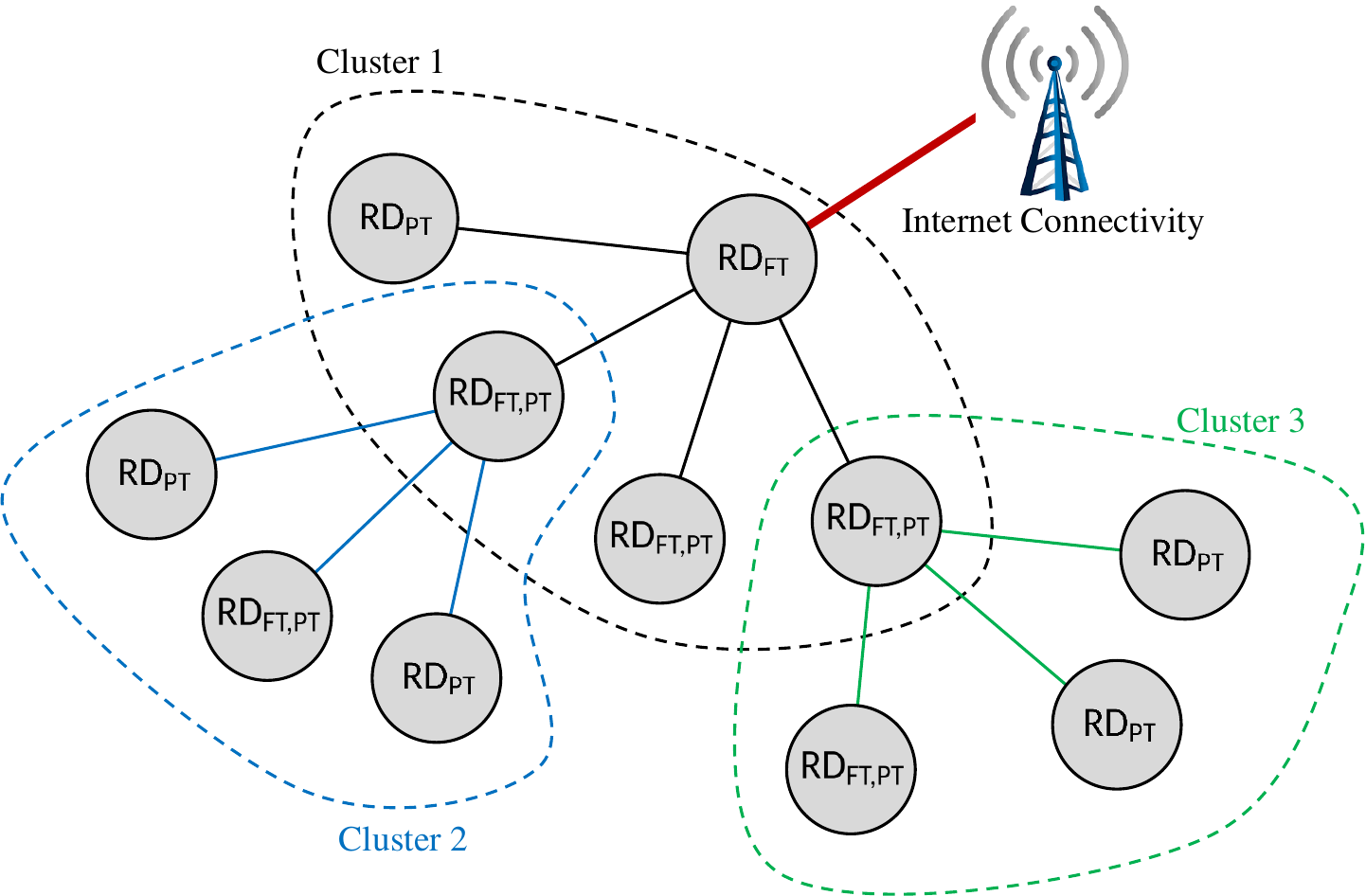}
\vspace{-4.5ex}
\caption{Illustration of a conventional multihop DECT-2020 NR Network containing multiple relay nodes, many leaf nodes and one gateway node that has internet connectivity multiple clusters.}
\label{fig:sys_NR}
\end{figure}

\begin{figure}[H]
    \centering
    \includegraphics[width=\columnwidth]{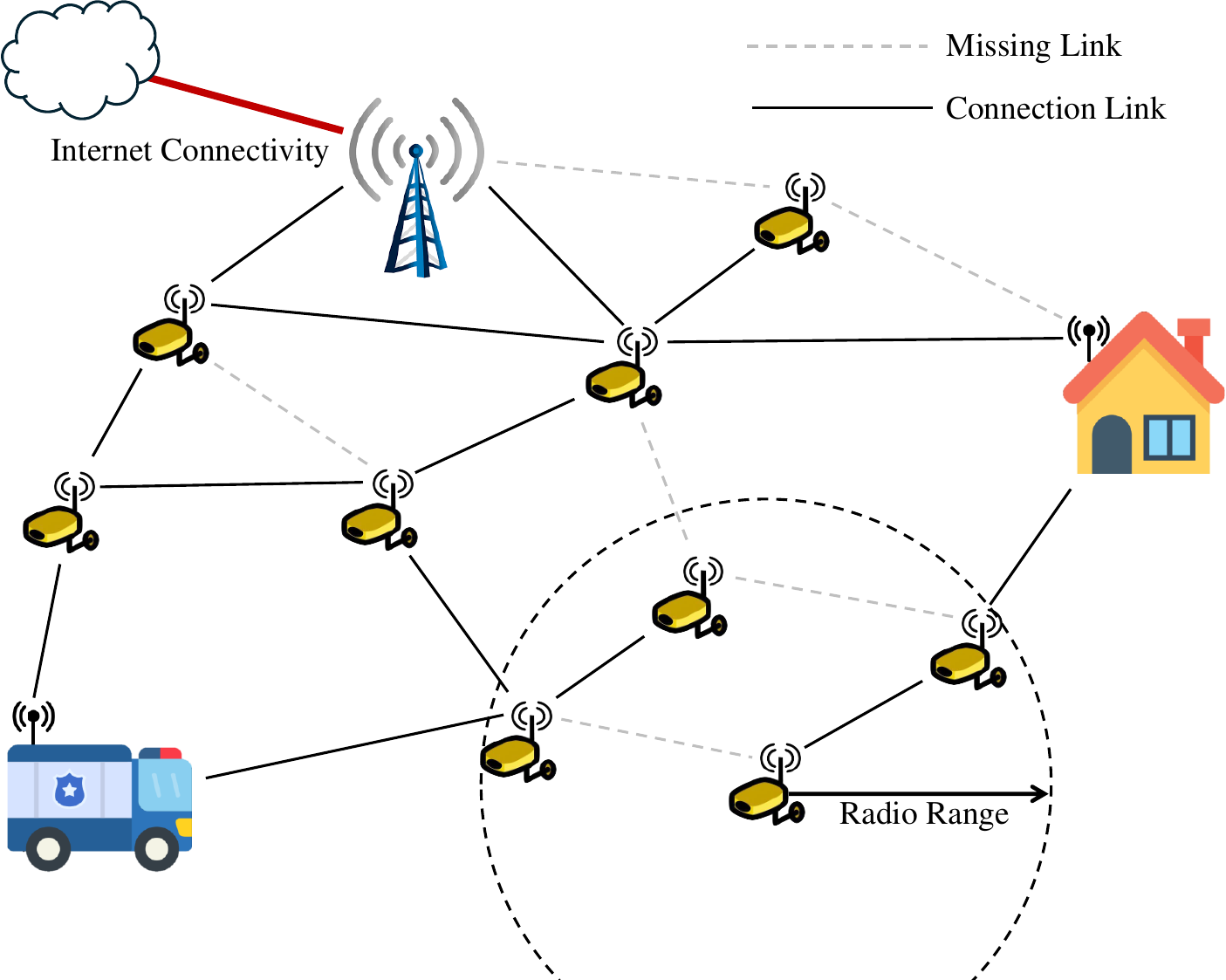}
    \vspace{-4.5ex}
    \caption{Illustration of the system scenario of a mesh network with multiple relay nodes and three gateways, where a link between nodes are displayed by solid lines and temporarily blocked links are illustrated by dashed lines, additionally displaying the radio range by the dashed circle.}
    \label{fig:sys}
\end{figure}

Consider a \ac{2D} DECT-2020 NR 5G mesh network as depicted in Fig.~\ref{fig:sys}, with a network density $\lambda$, containing $N$ nodes located at the coordinates $\bm{\Theta}=[\bm{\theta}_1,\cdots,\bm{\theta}_{N-1}]$.
It is assumed that only a subset of nodes are aware of their location, which are denoted as gateways and have their coordinates denoted by $\bm{A}=[\bm{a}_1,\cdots,\bm{a}_{N_a}]$, with $\bm{A}\subseteq\bm{\Theta}$ and $N_a>2$, where the nodes follow a tree structure and only some nodes can receive (PT), relay (FT), or do both (FT,PT)\footnote{Without loss of generality (w.l.g.), it can be assumed that the nodes are deployed as a mesh network, where all nodes are capable of the FT,PT NR+ mode.} as illustrated in Figure \ref{fig:sys_NR}. 

It can be assumed that two nodes $i$ and $j$ are mutually connected \ac{iff} the distance between them is less than a given threshold that we denote as the radio range, which we set as the unit circle such that the distance between two nodes $i$ and $j$ is given by $D_{i,j}=||\bm{\theta}_i-\bm{\theta}_j||_2\leq1$.
These type of \ac{2D} networks are commonly referred to as \acp{UDPN}, where w.l.g. all Euclidean distances can be assumed to be normalized to the corresponding communication threshold.
While the true distance between two nodes is defined as $D_{i,j}$, the multihop distance between them is defined by the sum of the distances of the different hops, given by
\begin{equation}
    \tilde{D}_{i,j}=\sum_{l=1}^{N_H}D_{l-1,l},
\end{equation}
where $N_H$ is the number of hops between the nodes $i$ and $j$, and $D_{l-1,l}$ is the distance between the nodes $l-1$ and $l$ in the $l$-th hop, with the total multihop distance $\tilde{D}_{i,j}\gg D_{i,j}$. 

For this article's specific scenario, it is assumed that a certain number of edges between nodes are missing. These can be the result of obstacles, such as buildings, trees, or large vehicles which block messages between nodes; or other factors like the devices' limited communications range or the presence of noise in the communication channel.
An example topology for the mesh network is depicted in Fig.~\ref{fig:sys}, where the nodes (circles) are connected by the edges (lines), with the missing edges shown as dashed lines.

\vspace{-2ex}

\vspace{-4ex}

\section{Proposed method}
\label{sec:prop}

\subsection{Network Discovery}
\subsubsection{Incomplete Network Discovery}

There are many \ac{SotA} methods for network discovery which can be used to find the routes between the nodes, such as the \ac{BFS} algorithm \cite{Bundy_1984}, which is a conventional method used to find the shortest path between two nodes in a network, or the \ac{AODV} routing scheme \cite{Perkins_1999} that floods the network to find the routes between the nodes, not saving the whole route but only the next hop to reach the destination.
However, in the context of real-world scenarios, the network discovery process is often incomplete due to factors like obstacles, limited communications range and channel noise.
Thus, again, as depicted in Fig.~\ref{fig:sys}, network discovery can be considered as the process of finding end-to-end routes between nodes, where some of the edges linking nodes are missing, leading to inefficient route observations and high number of hops between nodes.

To perform the network discovery process in the proposed scenario, we first perform a \ac{BFS} algorithm to find the routes between the nodes, based on the observed incomplete graph with the missing edges.
However, due to the temporarily blocked links, the \ac{BFS} algorithm will fail to report all the potential routes between the nodes, leading to an incomplete hop matrix $\bm{H}\in\mathbb{R}^{N_a\times N}$, where again, $N_a$ is the number of gateways and $N$ is the number of nodes in the network.
Thus, the next step after the first iteration of network discovery is to recover the missing entries of the hop matrix, followed by the discovery of the recovered and improved routes between the nodes in a second discovery process.
In contrast to the \ac{SotA} methods, the proposed method performs the network discovery process in a decentralized manner by only employing a portion of the total nodes, without flooding the whole network with routing messages, but rather by finding the routes between the nodes more efficiently.

\subsubsection{A Note on the State-of-the-Art}
\label{sec:SotA}

While the first iteration of the network discovery step discussed above yields an incomplete hop matrix and route table, conventional \ac{SotA} methods, such as \ac{BFS}, or \ac{AODV}, can also be used to find the routes between the nodes based on the observed incomplete graph with the missing edges. 
Thus, either a \ac{BFS} algorithm with a truncated adjacency matrix would be performed, or network flooding methods, such as \ac{AODV} Routing would be used to find the routes between the nodes, neglecting the temporarily unavailable edges.
In the proposed scenario, the \ac{SotA} method is chosen to be a second iteration of \ac{BFS} algorithm, which is performed on the incomplete network, neglecting the temporarily unavailable edges, leading to a complete but suboptimal route table and hop matrix.
Note, however, that in contrast to the decentralized efficient routing process proposed in this article, the \ac{SotA} method would flood the network with routing messages in a centralized manner, which can lead to a high overhead in the network, especially in the presence of a large number of nodes.

\subsection{Hop Recovery via Matrix Completion}

With the incomplete hop matrix in hand, matrix completion techniques such as OptSpace \cite{OptSpace} or Soft-Impute \cite{Yao_2019} can be used to recover the missing entries corresponding to the hop distances between the nodes.
Instead of using standard matrix completion techniques, however, we propose to use a discrete-aware matrix completion scheme \cite{Iimori_2020, Nic_Asilo_2024}, which is specifically designed for the case where the missing entries belong to a finite discrete alphabet set.
In our scenario, the finite discrete alphabet set is defined as $\mathcal{S}\triangleq\{1,2,\cdots,\text{max}(\bm{H})+1\}$, since any missing targets can be at most one more hop away from an existing route.
The alphabet set $\mathcal{S}$ is then used in the discrete-aware matrix completion scheme first proposed in \cite{Iimori_2020}\footnote{Note that the discrete-aware matrix completion scheme from \cite{Iimori_2020} is chosen over the slightly better performing variation of \cite{Nic_Asilo_2024}, due to its advantages in terms of computational complexity.}, to recover the missing hop distances by assuming that the missing entries of the matrix belong to the aforementioned set.

In view of the latter, the corresponding standard regularized optimization problem based on rank minimization \cite{RankEDM,Jain_2009,Tanner_2016} can be rewritten as
\vspace{-1ex}
\begin{equation}
\argmin_{\boldsymbol{H}\in \mathbb{R}^{N_a\times N}}\quad f(\boldsymbol{H})+\lambda g(\boldsymbol{H})+\zeta r(\boldsymbol{H}|p),
\end{equation}
where $g(\boldsymbol{H})$ denotes a typical low-rank regularizer, which is chosen to be the \acf{NN} that relaxes the problem and was shown to be a tight lower bound of the rank operator \cite{Recht_2010}, while the discrete-space regularizer $r(\boldsymbol{H}|p)$ defined as
\vspace{-1ex}
\begin{equation}
r(\boldsymbol{H}|p)\triangleq \sum_{k=1}^{|\mathcal{S}|}||\text{vec}_{\bar{\Omega}}(\boldsymbol{H})-a_k\boldsymbol{1}||_p,
\vspace{-1ex}
\end{equation}
where $\text{vec}_{\bar{\Omega}}(\boldsymbol{H})$ represents a vectorization of the matrix $\boldsymbol{H}$, where the entries are chosen corresponding to the given index set $\bar{\Omega}$, being the complementary set to the previously presented set $\Omega$ and $0\leq p$, which is chosen as $p=1$ according to \cite{Iimori_2020}.

Following \cite{Iimori_2020, Nic_Asilo_2024}, we point out that a \ac{PG} method can be used to solve the problem iteratively, containing the following steps:
\begin{eqnarray}
&\boldsymbol{Y}_t=(1+\gamma_t)\boldsymbol{H}_{t-1}+\gamma_t\boldsymbol{H}_{t-2},\label{eq:acc_old}&\\
&\boldsymbol{Z}_t=\text{prox}_{\zeta_r}(\boldsymbol{Y}_t),&\\
&\boldsymbol{H}_t=\text{SVT}_\lambda(P_{\bar{\Omega}}(\boldsymbol{Z}_t)+P_{\Omega}(\boldsymbol{O})),&
\end{eqnarray}
where the first step integrates a moment acceleration function, the second corresponds to the proximal operation on the discrete space regularizer, and the last step contains the \ac{PG} operation on the \ac{NN} regularizer $g(\boldsymbol{H})$, since it is known that \ac{SVT} can be used as a proximal minimizer for the \ac{NN} \cite{Cai_2010}.

Additionally, $P_{\Omega}(\cdot)$ indicates a mask operator, defined as 
\vspace{-1ex}
\begin{equation}
[P_{\Omega}(\boldsymbol{H})]_{i,j}=
\begin{cases}
[\boldsymbol{H}]_{i,j},& \text{if } (i,j) \in \Omega, \\
0,              & \text{otherwise},
\end{cases}
\vspace{-1ex}
\end{equation}
where $[\cdot]_{i,j}$ denotes the $(i,j)$-th element of a given matrix, and $\Omega$ denotes the index set of the observed elements.

\subsection{Privacy-preserving Multihop Localization}

After the recovery of the incomplete hop matrix, the next step is to obtain a complete routing table, which can be used to establish the final communication paths between the nodes.
To that end, we use the multihop localization technique proposed in \cite{Rahmatollahi_2011}, which allows the nodes to estimate their positions based on the hop distances to a set of known gateways, without sharing their positions or the distances themselves with their neighbors, thus preserving privacy.
The following steps, revisited from \cite{Rahmatollahi_2011} are performed to estimate the position of the nodes:

\subsubsection{Vicinity Discovery}

The first step of the scheme is to perform vicinity discovery, which implies that the nodes need to find their neighbors in the network, which can be used to estimate the distances to the gateways.
Thus, the node can broadcast a discovery message to its neighbors within radio range, which can then respond with their own discovery messages\footnote{Note that to preserve privacy, since the nodes want to protect their positions they can reply with a random delay to the discovery messages, such that the position cannot be inferred from the time of arrival of the messages.}, containing their own identities.

As a result, each node can obtain a local estimate of the network density $\tilde{\lambda}$ by counting the number of neighbors that respond to its discovery message.
By obtaining an estimate of the network density as shown in \cite[Fig. 3]{Rahmatollahi_2011}, the average hop length $\bar{d}_{BFS}$ in a \ac{BFS} discovery can be found, which is needed to estimate the distances to the gateways.

\subsubsection{Gateway Discovery}

In the \ac{SotA} localization was an addition to the routing, performed via a complete \ac{BFS} discovery, while in our proposed scheme, the localization is an active step using the completed hop matrix, in order to find all missing routes.
Thus, the first step for the gateway discovery includes, after the actual selection of the gateways\footnote{W.l.g. in the proposed scenario, the gateway nodes can serve as a connection to the internet, those positions are commonly known and can thus be chosen as known reference points for the localization.} to perform a network discovery, as it has been done in Section \ref{sec:prop}, to obtain the number of hops $N_H$ between the gateways and the target nodes.
Finally, after all parameters are obtained, following \cite[Theorem 2]{Rahmatollahi_2011}, first presented in \cite{Rahmatollahi_2012}, an upper bound on the target nodes and the gateways can be obtained and used to estimate the position of the nodes in the next step.

\subsubsection{Multihop Localization}

For the process of localization, following the procedure in \cite{Rahmatollahi_2011}, the conventional \ac{CSDP} algorithm, based on the well-known SDP algorithm presented in \cite{So2007}, augmented by an additional constraint is used to localize the target $\bm{\theta}_1$, $i.e.$,
\vspace{-1ex}
\begin{subequations}
    \begin{align}
        & \underset{\bm{\theta}_1\in\mathbb{R}^{2\times1}}{\text{maximize}} \quad 0 \\
        & \text{subject to} \quad \mathbf{Z}_{1:2,1:2} = \mathbf{I}_2, \\
        & \phantom{\text{subject to}} \quad 0 < \operatorname{tr}(\mathbf{C}_A \mathbf{Z}) \leq \tilde{D}_{1i}, \forall i \in \bm{A}\label{eq:CSDP}\\
        & \phantom{\text{subject to}} \quad \mathbf{Z} \succeq 0,
    \end{align}
\end{subequations}
\vspace{-2ex}
where $\mathbf{C}_A$ is defined as
\begin{equation}
    \mathbf{C}_A \triangleq 
    \begin{bmatrix}
        \mathbf{a}_i \\ -1
    \end{bmatrix}
    \begin{bmatrix}
        \mathbf{a}_i^\intercal & -1
    \end{bmatrix},
\end{equation}
and $\mathbf{Z}_{1:2,1:2}$ is the $2 \times 2$ first upper left sub-matrix of the $3 \times 3$ matrix $\mathbf{Z}$, which contains the estimated node position $\hat{\theta}_1$ and is defined as
\begin{equation}
    \mathbf{Z} \triangleq 
    \begin{bmatrix}
        \mathbf{I}_2 & \hat{\theta}_1 \\
        \hat{\theta}_1^\intercal & y
    \end{bmatrix},
\end{equation}
where $\mathbf{I}_2$ denotes a $2 \times 2$ identity matrix and $y = \hat{\theta}_1^\intercal \hat{\theta}_1$.

Notice that since the nodes cannot have the same location, the lower bound on the node-to-gateway distance in equation \eqref{eq:CSDP} is $0$, while the upper bound is the estimated distance $\tilde{D}_{0i}$.

\subsection{Local Discovery}

Finally, a local discovery step is performed to complete the hop matrix with real information, finding the final route table.
We make use of the known radio range and the previously estimated positions, to define which node should perform the local discovery.
%
%
Therefore, the gateways can first find all nodes that are within the radio range of the target by using the knowledge of the positions of the nodes\footnote{Note that if there are no nodes within the radio range of the target, it is either unreachable, or the localization step yielded poor estimates, which results in a local discovery with the nodes closest to the target.}, and then sort the nodes according to the already measured hops required to reach them from the perspective of the gateways, since every node can only be at most one hop away from a connection that the gateways already acquired. 
Thus, after defining the nodes that should perform the local discovery, these nodes can perform a local network flooding discovery to obtain the final hop matrix, and establish the final routes between the nodes, hence yielding a completed hop matrix, as well as the final route table.
This step does not only yield the final routes between the nodes, but also lowers the complexity and messages required to establish the routes, since only a few nodes need to perform the local discovery in a decentralized, instead of flooding the whole network with routing messages, which can be performed iteratively whenever necessary.

\vspace{-1ex}

\section{Performance Evaluation}
\label{sec:res}
\vspace{-1ex}

In this section we provide simulation results illustrating the performance of the proposed routing technique compared to the \ac{SotA}.
The performance metric of choice is the average number of hops, which can be calculated by the average of the corresponding hop matrix $\bm{H}$ that represents the length of the obtained routes, as a function of the network density $\lambda$, as well as the amount of incompleteness, $i.e.$, the percentage of missing edges.
Each point in the figures is obtained by averaging $K=10^3$ Monte-Carlo realizations, using a network with $N=100$ nodes, $N_a=10$ gateways, and a radio range of $1$ meter, with the region of interest being $[10\text{m};10\text{m}]$.
%

\begin{figure}[H]
\centering
\includegraphics[width=\columnwidth]{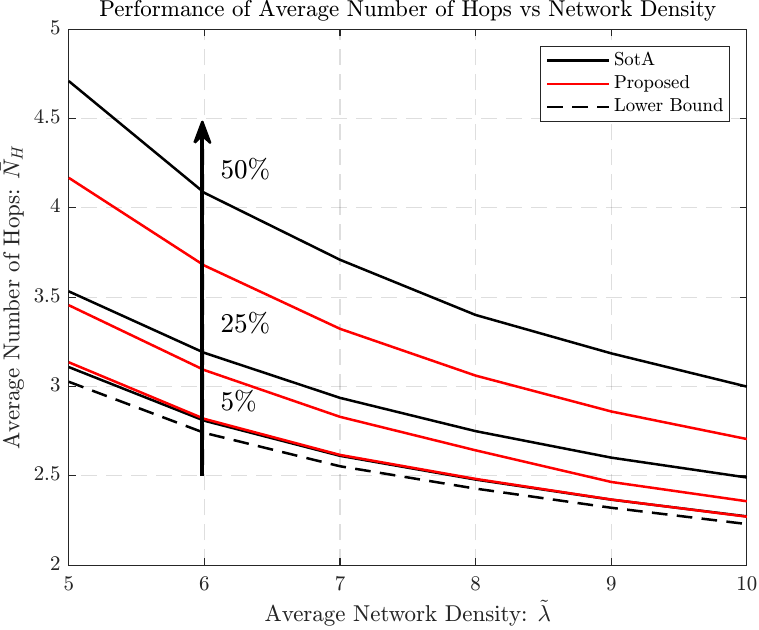}
\vspace{-5ex}
\caption{Average number of hops over the network density for different levels of missing edges.}
\label{fig:Avg_Dens}
\end{figure}

The results in Fig.~\ref{fig:Avg_Dens} show that in terms of average hops, the proposed technique outperforms the \ac{SotA} over all network densities, while maintaining a performance close to the optimal bound with fewer missing edges.

\begin{figure}[H]
\centering
\includegraphics[width=\columnwidth]{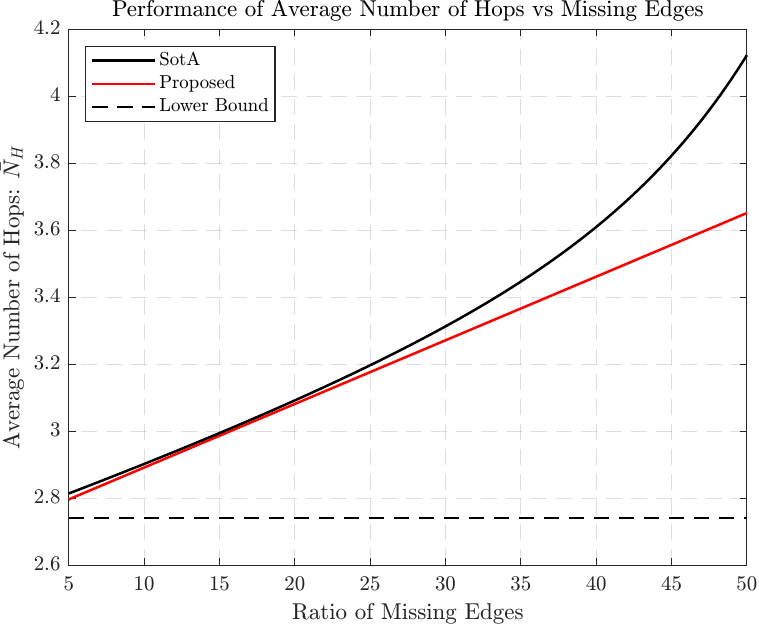}
\vspace{-5ex}
\caption{Average number of hops over the amount of incompleteness for $\tilde{\lambda}=6$.}
\label{fig:Avg_IO}
\end{figure}

This bound can be found by performing the \ac{BFS} algorithm without missing links.
Our second set of results in Fig.~\ref{fig:Avg_IO} show that the proposed technique is more robust to incompleteness in the hop matrix than the \ac{SotA}, as it maintains a lower average number of hops between nodes even as the number of missing edges increases.
%
    

\vspace{-1ex}

\vspace{-4ex}
\newpage

\vspace{-2ex}

In a follow-up work, more realistic models can be implemented, to measure data with real-life measurements and cost of each path, $e.g.$, in terms of transmission power or delay, such that routes that consists of the same number of hops can vary by their total length, delay, or power consumption, which can be used to optimize the routing process further.
Additionally, it can be noted that by leveraging localization techniques with a higher accuracy, the proposed routing scheme can be further improved, however, it would come at the cost of a potentially higher computational complexity, and a larger amount of position information being leaked. 

\vspace{-1ex}
\section{Conclusion}
\vspace{-1ex}
We proposed a novel routing scheme suitable for ad-hoc DECT 2020 NR systems, which combines network discovery and route completion techniques to establish communication paths efficiently and reliably, where links can temporarily be unavailable. 
To that extend, a network discovery technique was first introduced, which uses an incomplete breadth-first search algorithm to discover the network topology, due to missing edges.
Then, a discrete-aware matrix completion technique was used to recover the missing hop distances, which were used to estimate the positions of the nodes via a multihop localization technique, while preserving privacy.
Finally, a local discovery technique was used to obtain the final hop matrix and establish the final routes between the nodes by leveraging the known radio range and the previously estimated positions of the nodes.
Simulation results illustrate the good performance of the proposed technique in terms of the average number of hops as a function of the network density and the amount of incompleteness in the hop matrix, compared to the \ac{SotA}, specifically, being more robust to incompleteness  due to the matrix completion and multihop localization techniques used in the proposed scheme.
Additionally, in contrast to the \ac{SotA} which always requires a complete network discovery to find the routes between the nodes, the proposed scheme allows the network to iteratively run decentralized routing requests at any time.

\vspace{-1ex}
\section*{Acknowledgments}
\vspace{-1ex}
This work was conducted as part of the "5G-HyprMesh" project with code 01MO24001C, funded by the Bundesamt für Sicherheit in der Informationstechnik (BSI).




\vspace{-1.5ex}

\end{document}